\documentclass{mem}
\usepackage{natbib}\usepackage{txfonts}\usepackage{balance}
\usepackage{graphicx}
\usepackage{txfonts}
\usepackage[a4paper]{hyperref}
\idline{79}{3}
\newcommand{\omc}{\mbox{$\omega$ Cen~}} 

\begin{document}
\def\teff{$T\rm_{eff }$}
\def\kms{$\mathrm {km s}^{-1}$}

\title{On the radial distribution of white dwarfs in the Galactic globular cluster \omc}

   \subtitle{}

\author{
A. \, Calamida\inst{1}
\and  C. E. \, Corsi\inst{1}
\and  G. \, Bono\inst{1}
\and  P. B. \,Stetson\inst{2}
\and  P. G. \, Prada Moroni\inst{3}
\and  S. \, Degl'Innocenti\inst{3}
\and  I. \, Ferraro\inst{1}
\and  G. \, Iannicola\inst{1}
\and  D. \, Koester\inst{4}
\and  L. \, Pulone\inst{1}
\and  M. \, Monelli\inst{5}
\and  P. \, Amico\inst{6}
\and  R. \, Buonanno\inst{7}
\and  L. \, M. Freyhammer\inst{8}
\and  E. \, Marchetti\inst{6}
\and  M. \, Nonino\inst{9} 
\and  M. \, Romaniello\inst{6}
        }


\institute{
INAF-OAR, via Frascati 33, 00040 Monte Porzio Catone, Italy
\email{calamida@mporzio.astro.it}
\and
DAO, HIA-NRC, 5071 West Saanich Road, Victoria, BC V9E~2E7, Canada
\and
Dipartimento di Fisica "E. Fermi", Univ. Pisa, Largo B. Pontecorvo 2, 56127 Pisa, Italy
\and
University of Kiel, 24098 Kiel, Germany
\and
IAC, Calle Via Lactea, E38200 La Laguna, Tenerife, Spain
\and
ESO, Karl-Schwarzschild-Str. 2, D-85748 Garching, Germany
\and
Universit\`a di Roma Tor Vergata, via della Ricerca Scientifica 1, 00133 Rome, Italy
\and
University of Central Lancashire, Preston PR1 2HE, UK
\and
INAF-OAT, via G. B. Tiepolo 11, 40131 Trieste, Italy}

\authorrunning{Calamida et al.}

\titlerunning{Radial distributions of \omc white dwarfs}

\abstract{
We present deep and accurate photometry ($F435W$, $F625W$, $F658N$)
of the Galactic Globular Cluster \omc collected with the Advanced Camera for
Surveys (ACS) on board the Hubble Space Telescope (HST). We identified
$\approx 6,500$ white dwarf (WD) candidates and compared their radial distribution 
with that of Main Sequence (MS) stars. We found a mild evidence that young 
WDs ( $0.1 \lesssim t \lesssim 0.6$ Gyr) are less centrally concentrated when 
compared to MS stars in the magnitude range 25$ \le F435W \le $26.5.  
}
\maketitle{}

\section{Introduction}
The observation of white dwarfs (WDs) in Galactic Globular clusters (GGCs)
presents some undisputed advantages: {\it 1)} - They are located at the same
distance and have approximately the same reddening. Moreover, the colors of WDs
are, at all luminosities, systematically bluer than those of Main Sequence (MS)
stars. This means that we can use a Color-Magnitude Diagram (CMD) instead of a 
color-color plane to identify cluster WDs. Therefore, WDs in GGCs are not
affected by color degeneracy with MS stars such as field WDs (Hansen \& Liebert 2003);
{\it 2)} - According to current evolutionary predictions in a
GGC with an age of $\approx$ 12 Gyr and by assuming a Salpeter-like initial 
mass function ($\alpha=2.35$) the number of WDs is three orders of magnitude 
larger than the number of Horizontal Branch (HB) stars (Brocato et al.\ 1999). 
This, together with the high stellar concentration of GGCs, implies that 
the expected local density of WDs in these systems is several 
orders of magnitude higher than the local density in the Galactic field;\\
{\it 3)} - We can trace back the evolutionary properties of the progenitors
of cluster WDs, since both the chemical composition and the age of GGCs 
are well known (Kalirai et al.\ 2007).
However, the observation of WDs in GGCs also presents a drawback: 
cluster WDs are faint stars and are severely affected by crowding problems.
Therefore, both photometry and medium resolution spectroscopy are difficult 
even for 8m class telescopes (Moehler et al.\ 2004).

In a previous investigation based on three out of nine pointings of the 
Advanced Camera for Surveys (ACS) on board the Hubble Space Telescope (HST),
we have already addressed the properties of WDs in \omc (Monelli et al 2005, 
hereafter MO05). Furthermore, Calamida et al.\ (2007b, hereafter CA07), 
based on eight out of the nine ACS pointings, identified $\approx 6,500$ WD 
candidates. They found that the ratio of WD and MS star counts is at least a 
factor of two larger than the ratio of CO-core WD cooling times and MS lifetimes. 
The presence of He-core WDs might explain the observed star counts, 
and the required fraction of He-core WDs ranges from 10\% to 80\%,
depending on their mean mass. 
We now adopt our sample of WD candidates to 
investigate their radial distribution. Recent observations of WDs in
M4 and NGC~6397 (Davis et al.\ 2006;\ 2007) showed that young WDs ($t \le 1$ Gyr)
are less centrally concentrated than either older WDs or progenitor MS stars.
Davis et al.\ suggest that these WDs are born with a natal kick, starting 
their life with a larger velocity dispersion when compared to the velocity 
of neighboring stars. Therefore, within a short time scale, the young WDs 
would acquire a more extended radial distribution. In order to explain this 
evidence, the authors suggest that these WDs have acquired a kick during their 
asymptotic giant branch phase caused by a slightly asymmetric 
mass loss (Spruit 1998; Heyl 2007). 

\begin{figure*}
\resizebox{\hsize}{!}{\includegraphics[clip=true]{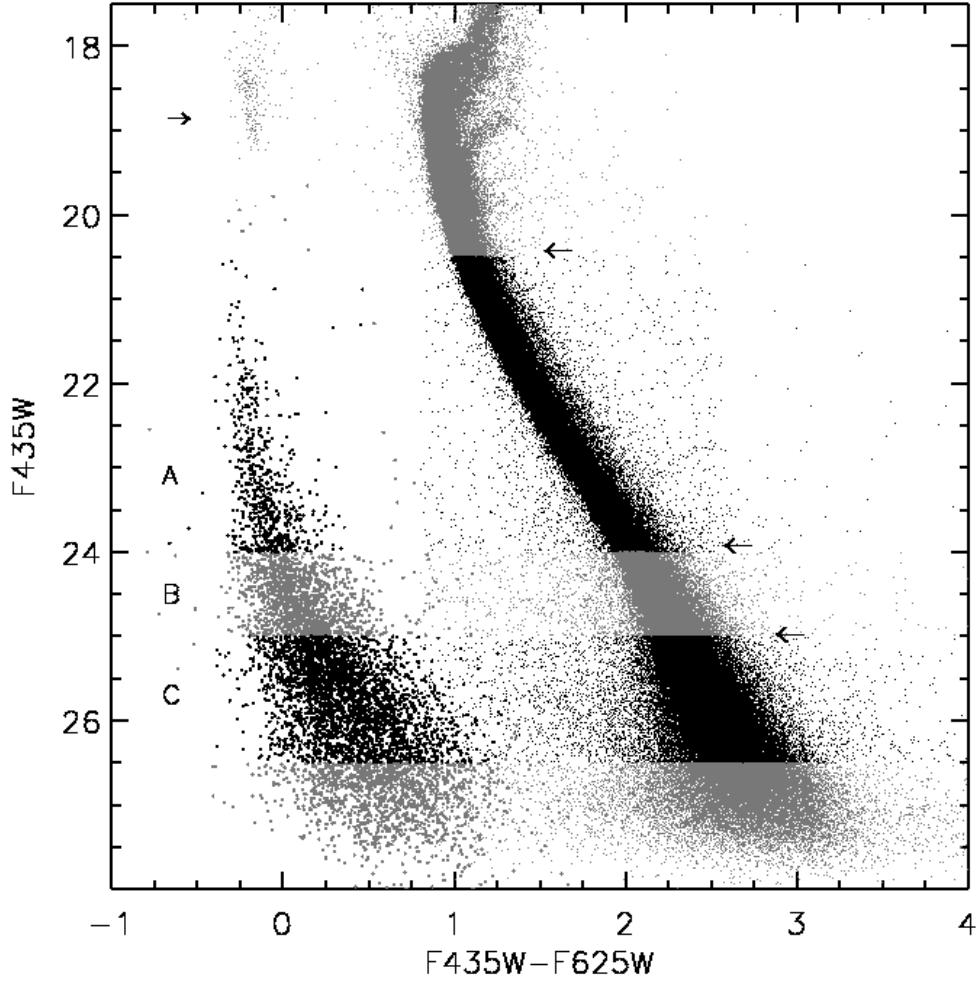}}
\caption{\footnotesize{$F435W,\ F435W-F625W$ CMD based on deep images collected with 
ACS@HST and reduced with ROMAFOT for the WD cooling sequence and shallow and deep 
images reduced with ALLFRAME for the MS. The different grey levels indicate the WD 
and MS samples selected for the radial distributions. The right arrows mark the 
bright magnitude levels of the selected MS samples. The left arrow indicates
the EHB stars. The letters mark the selections along the WD cooling sequence.}
}
\label{fig1}
\end{figure*}

\section{Observations and data reduction}
Multiband ($F435W$, $F625W$, $F658N$) photometric data collected 
with the ACS on board the HST were retrieved from the HST archive. 
The current data set includes eight out of the nine pointings located 
across the cluster center that have already been discussed by Castellani 
et al.\ (2007, see their Fig.~1). The central pointing of the 
$3\times 3$ mosaic was omitted due to the severe crowding of the 
innermost regions. For each pointing four images in three different bands 
were acquired. The $F435W$- and $F625W$-band data consist of one shallow 
($8$s) and three deep ($340$s each) exposures, while the $F658N$-band data 
consist of four exposures of $440$s each per field. The raw frames were 
pre-reduced by the standard HST pipeline. The photometric catalogs (96) were 
rescaled to a common geometrical system with {\tt DAOMATCH/DAOMASTER}, and
the entire set of images was then re-reduced simultaneously with 
{\tt DAOPHOT~IV/ALLFRAME}. The final catalog includes 
more than one million stars having at least one measurement in two 
different photometric bands. The photometry was kept in the Vega 
system following the prescriptions suggested by Sirianni et al.\ (2005).

We then selected, from the final ALLFRAME catalog, all the stars systematically bluer 
than MS stars and fainter than extreme HB (EHB) stars ($F435W \lesssim 20$), ending
up with a sample of $\approx$ 60,000 stars. The photometry of these stars 
was performed once again using ROMAFOT (Buonanno  \& Iannicola 1989), but only 
for the deep exposures, namely three $F435W$, three $F625W$, 
and four $F658N$ images per pointing. Individual stars have been interactively checked 
in every image, and the WD candidates were measured either as isolated stars 
or together with neighbor stars in simultaneous joint solutions. Note that most  
of the neighbor stars located close to WD candidates are truly MS stars, 
i.e., they did not belong to the original sample of stars located 
on the blue side of the MS. A significant fraction of the originally selected 
detections turned out to be either cosmic rays or spurious detections 
close to saturated stars, or detections too faint 
to be reliably measured on individual images. 
Fig.~1 shows the $F435W,\ F435W-F625W$ CMD based on the ROMAFOT photometry for 
the refined sample of WD candidates and on the ALLFRAME photometry for the 
MS, the sub-giant branch and the EHB stars. Data plotted in this figure show 
that the cluster WD candidates ($\sim 6500$) are distributed along a well 
defined star sequence fainter than EHB stars and systematically bluer 
than MS stars (MO05). To our knowledge this is the largest sample 
of cluster WD candidates ever detected (CA07).

\section{Radial distributions}
We adopted our sample of WD candidates to investigate the radial distribution of 
these stars in \omc.
We selected WDs in three magnitude bins, namely $20.5 \le F435W \le 24$ 
(A, see Fig.~1),  $24 \le F435W \le 25$ (B), and $25 \le F435W \le 26.5$ (C).
In order to estimate the cooling times of these WD samples, we adopted a 
predicted cooling sequence for a CO-core and H-rich envelope WD ($M = 0.5M_{\odot}$) 
by  Althaus, \& Benvenuto (1998). The theoretical predictions were transformed 
into the observational plane by adopting the pure H atmosphere model computed 
by Koester, \& Wolff (2000) and by Koester et al.\ (2005, for more details see CA07).
The corresponding cooling times are: $t \lesssim 20$ Myr (A), $t \lesssim 120$ Myr (B),
and $t \lesssim 570$ Myr (C). 
  
In order to compare the radial distributions of WDs with those of MS stars, we 
selected MS stars in the same three magnitude bins (see Fig.~1). Note that the 
selection of MS stars is based on the ALLFRAME catalog, while the selection of WDs 
is based on the ROMAFOT catalog. For the former data set the completeness along the 
MS is $\approx$ 80\% at $F435W$ = 24. In order to have  approximately the same 
completeness for WDs and MS stars, we compared the radial distributions of stars 
in the same magnitude bins.
\begin{figure*}[t!]
\resizebox{\hsize}{!}{\includegraphics[clip=true]{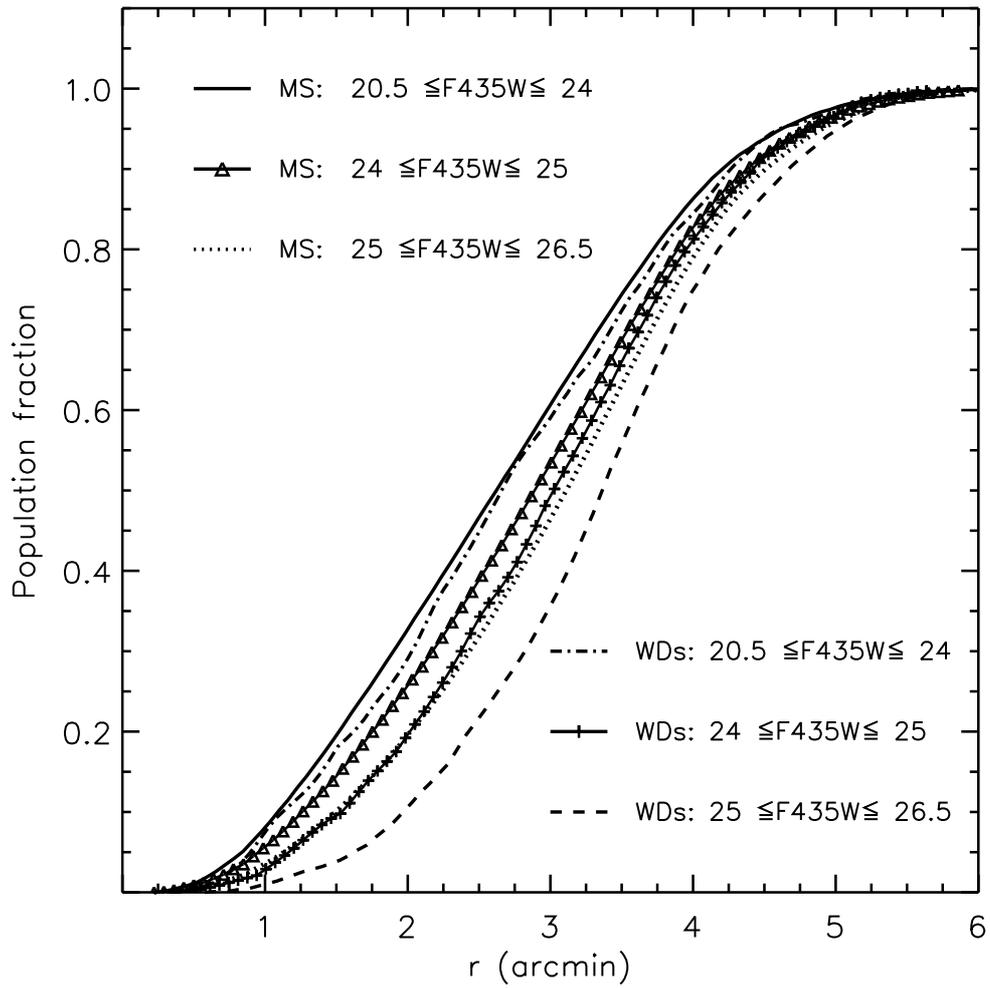}}
\caption{\footnotesize{Cumulative radial distributions of WD and MS stars.
The different samples are shown with different lines labeled on the figure.}
}
\label{fig2}
\end{figure*}

Fig.~2 shows the six cumulative radial distributions. 
The WD and MS star radial distributions are in agreement, within the uncertainties,
for the bright magnitude bin (20.5 $\le F435W \le$ 24). On the other hand, the WD radial distribution appears to be 
less centrally concentrated compared to the MS profile in the case of the intermediate 
(24 $\le F435W \le$ 25) and of the faint (25 $\le F435W \le$ 26.5) magnitude bins. 
The difference between the three MS radial distributions could be due to completeness 
problems (the crowding would affect more the fainter stars) or to the presence of 
mass segregation. Ferraro et al.\ (2006), based on ACS@HST and WFI@2.2m photometric data 
of \omc, showed that the Blue Stragglers radial distribution do not differ from 
the red giant branch and HB distributions, up to a distance of 20' from the cluster center. 
This evidence would suggest that \omc is not a fully relaxed stellar system. However, 
the relaxation time at the core radius is $t_{r_c} \sim 5.4$ Gyr, much 
shorter that the cluster age ($t \approx 12-13$ Gyr), and some mass segregation 
should be observable. In order to assess if mass segregation is really present 
in \omc, we should first estimate the completeness along the MS up to $F435W \sim 26.5$.
On the other hand, there is a mild evidence of a deficiency of WDs in the cluster 
center ($r \lesssim$ 6'), as shown by the discrepancy between WD and MS star radial 
distributions in two magnitude bins (see Fig.~2). 

\section{Conclusions}
We adopted our sample of $\approx 6,500$ WD candidates to compare their radial 
distribution with that of MS stars. We selected WDs and MS stars in three $F435W$-band 
magnitude bins in order to have approximately the same completness level. 
We found a mild evidence that young WDs ($0.1 \lesssim t \lesssim 0.6$ Gyr) are 
less centrally concentrated when compared to MS stars in the magnitude range 
25 $\le F435W \le$ 26.5. This evidence would support the results of 
Davis et al.\ (2006; 2007) who found that young WDs in NGC~6397 and M4 have an 
extended radial distribution when compared to the most massive MS stars in the clusters.


%


\bibliographystyle{aa}

\end{document}